\documentclass[12pt,preprint,notoc,nohyper]{QCEMDA} 

\newcommand\fverb{\setbox\pippobox=\hbox\bgroup\verb}
\newcommand\fverbdo{\egroup\medskip\noindent%
            \fbox{\unhbox\pippobox}\ }
\newcommand\fverbit{\egroup\item[\fbox{\unhbox\pippobox}]}
\newbox\pippobox

\title{Quantum tunneling from scalar fields in rotating black strings}

\author{H. Gohar and K. Saifullah  \\

Department of Mathematics, Quaid-i-Azam University, Islamabad,
Pakistan \\

Electronic address: \email{saifullah@qau.edu.pk}}

\preprint{}  

\abstract{Using the Hamilton-Jacobi method of quantum tunneling and
complex path integration, we study Hawking radiation of scalar
particles from rotating black strings. We discuss tunneling of both
charged and uncharged scalar particles from the event horizons. For
this purpose, we use the Klein-Gordon equation and find the
tunneling probability of outgoing scalar particles. The procedure
gives Hawking temperature for rotating charged black strings as
well.}

\begin{document}

\section{Introduction}

Classically, black holes are perfect absorbers and do not radiate
any particles. However, this picture changes if we incorporate quantum effects into the theory of black holes. 
In the last forty years many advancements in the field of black hole physics came about as a result of the interplay
between classical thermodynamics and quantum properties of black
holes. In 1970's Bakenstein related the properties of black holes
with the laws of thermodynamics \cite{cite1}. Soon after this,
Hawking showed that quantum mechanically black holes radiate
particles \cite{1, 2, new2}. This discovery was very important
because it gave a new perspective to the quantum theory of gravity.
Following this researchers showed a great interest in the field of
black hole physics and used different methods to investigate thermal radiations from black holes.

In 1990's Kraus and Wilczek \cite{3, 4} developed the technique of
studying Hawking radiation as a phenomenon of quantum tunneling. In
this semi-classical approach, the imaginary part of the classical
action is calculated for outgoing trajectories across the horizon. 
By using WKB approximation, the tunneling probability for a
classically forbidden trajectory coming from inside to outside the
horizon is calculated. Then using the Boltzmann factor \cite{5, 6, DJ1} we can write
Hawking temperature for the black hole.

Originally the tunneling method was applied to the Schwarzschild
black hole \cite{3}. However, this proved to be a powerful method
and has been applied to a variety of black configurations
(\cite{new3}-\cite{CQGRev}) since then. There are two different ways to
calculate the imaginary part of the classical action for the
emitted particle: the null geodesic method and the Hamilton-Jacobi
ansatz. The first one was used by Parikh and Wilczek \cite{5a},
which followed from the work of Kraus and Wilczek, and the second
one, which is the extension of the complex path analysis \cite{5, 6}
has been used by different authors. Recently these radiations have been studied for charged black holes in string theory \cite{K}, squashed Kaluza-Klein black hole \cite{MU}, black holes in Einstein-Yang-Mills theory \cite{GD}, slowly rotating
Kerr-Newman black hole \cite{DLJC} and dilatonic black holes \cite{hu2, GM, GS}. These methods have also been applied to higher dimensional black holes \cite{H, WP, BCK}. 

The hoop conjecture in the theory of gravitational collapse excluded the formation of black holes for all objects that are not spherically symmetric. However this conjecture was formulated for spacetimes with zero cosmological constant and in the presence of negative cosmological constant one can expect black configurations with other symmetries as well, for example, those with cylindrical symmetry. After the pioneering work on cylindrically symmetric black holes \cite{16, 17, 18} a considerable amount of work has been done on their thermodynamical and other properties \cite{19, 20}. These black holes are also significant because they lead to the study of axially symmetric black holes, and because of their interaction with gravitational waves. Because of the negative cosmological constant these represent anti de Sitter spacetimes. 

In this paper we use the Hamilton-Jacobi ansatz to study tunneling
of scalar particles from cylindrically symmetric black holes, or
black strings \cite{16, 17}. In this method WKB approximation is used to solve a wave equation. For the present study we will solve the Klein-Gordon equation both for the uncharged as well as the charged cases. A solution is assumed by taking into account the symmetries of the background spacetime. Substituting this \textit{ansatz} the equation is integrated using complex path integration around the black hole horizon. The method has been shown to calculate the correct Hawking temperature for various black holes. Here we recover the correct value of Hawking
temperature for rotating black strings. The paper is organised as
follows. In the next section we explain the metric describing
rotating black strings. In Sections 3 and 4 we study tunneling of
uncharged and charged scalars, respectively, from these objects. At
the end we give a brief Conclusion.

\section{Rotating black strings}

We consider Einstein-Hilbert action in four dimensions with a
cosmological constant in the presence of the electromagnetic field.
Solving the Einstein-Maxwell equations for a charged rotating
cylindrically symmetric spacetime gives \cite{18, 19}
\begin{equation}
ds^{2}=-F(r)dt^{2}+\frac{dr^{2}}{N(r)}-H(r)dtd\theta +K(r)d\theta
^{2}+L(r) dz^{2},  \label{136}
\end{equation}%
where
\begin{eqnarray}
F(r) &=&\alpha ^{2}r^{2}-\frac{2G(M+\Omega )}{\alpha r}+\frac{4GQ^{2}}{%
\alpha ^{2}r^{2}},  \label{137} \\
N(r) &=&\alpha ^{2}r^{2}-\frac{2G(3\Omega -M)}{\alpha r}+\left(\frac{3\Omega -M}{%
\Omega +M}\right) \left(\frac{4GQ^{2}}{\alpha ^{2}r^{2}}\right),  \label{138} \\
H(r) &=&\frac{16GJ}{3\alpha r}\left( 1-\frac{2Q^{2}}{\left( \Omega
+M\right)
\alpha r}\right) ,  \label{139} \\
K(r) &=&r^{2}+\frac{4G(M-\Omega )}{\alpha ^{3}r}\left( 1-\frac{2Q^{2}}{%
\left( \Omega +M\right) \alpha r}\right) ,  \label{140} \\
L(r) &=&\alpha ^{2}r^{2}.
\end{eqnarray}%
Here $M$ and $Q$ are the ADM mass and charge of the black string,
$J$ is the angular momentum and $\Omega =\sqrt{M^{2}-8J^{2}\alpha
^{2}/9}$, where $\alpha^{2}=-\Lambda /3$, $\Lambda$ being the
cosmological constant. We can write Eq.(\ref{136}) in another form
\cite{18}
\begin{equation}
ds^{2}=-N^{0^{2}}dt^{2}+R^{2}\left( N^{\phi }dt+d\theta \right) ^{2}+\frac{%
dr^{2}}{g(r)}+e^{-4\phi }dz^{2},
\end{equation} 
where 
\begin{eqnarray*}
N^{0^{2}} &=&\left( \gamma ^{2}-\frac{\omega ^{2}}{\alpha
^{2}}\right)
^{2}\left( \alpha ^{2}r^{2}-\frac{b}{\alpha r}+\frac{c^{2}}{\alpha ^{2}r^{2}}
\right) \frac{r^{2}}{R^{2}}, \\
N^{\phi } &=&-\frac{\gamma \omega }{\alpha ^{2}R^{2}}\left( \frac{b}{\alpha r
}-\frac{c^{2}}{\alpha^2 r^2}\right) , \\
R^{2} &=&\gamma ^{2}r^{2}-\frac{\omega ^{2}}{\alpha ^{4}}\left(
\alpha
^{2}r^{2}-\frac{b}{\alpha r}+\frac{c^{2}}{\alpha ^{2}r^{2}}\right) , \\
g(r) &=&\left( \alpha ^{2}r^{2}-\frac{b}{\alpha
r}+\frac{c^{2}}{\alpha
^{2}r^{2}}\right) , \\
e^{-4\phi } &=&\alpha ^{2}r^{2},
\end{eqnarray*}%
and
\begin{eqnarray*}
b &=&4M\left( 1-\frac{3a^{2}\alpha ^{2}}{2}\right) , \\
c^{2} &=&4Q^2\left( \frac{1-3a^{2}\alpha
^{2}/2}{1-a^{2}\alpha ^{2}/2}\right) .
\end{eqnarray*}%
\newline
Here $N^{0^{2}}$ and $N^{\phi }$ are the lapse and shift functions
and $a$ is the rotation parameter such that $a=J/M$. Further,
$\gamma ^{2}$ and $\omega ^{2}/\alpha ^{2}$ are defind as
\begin{eqnarray*}
\gamma ^{2} &=&\frac{2GM}{b}+\frac{2G}{b}\sqrt{M^{2}-\frac{8J\alpha
^{2}}{9}} :\frac{\omega ^{2}}{\alpha
^{2}}=\frac{4GM}{b}-\frac{4G}{b}\sqrt{M^{2}-\frac{
8J\alpha ^{2}}{9}}, \\
\gamma ^{2} &=&\frac{2GM}{b}-\frac{2G}{b}\sqrt{M^{2}-\frac{8J\alpha
^{2}}{9}} :\frac{\omega ^{2}}{\alpha
^{2}}=\frac{4GM}{b}+\frac{4G}{b}\sqrt{M^{2}-\frac{ 8J\alpha
^{2}}{9}},
\end{eqnarray*}
or 
\begin{eqnarray}
\gamma &=&\sqrt{\frac{1-\frac{a^{2}\alpha
^{2}}{2}}{1-\frac{3a^{2}\alpha ^{2}
}{2}}},  \label{2aa} \\
\omega &=&\frac{a\alpha ^{2}}{\sqrt{1-\frac{3a^{2}\alpha ^{2}}{2}}}.
\label{2aaa}
\end{eqnarray}%
The line charge density along the $z$-line is given by
\begin{equation}
Q=\frac{Q_{z}}{\Delta z}=\gamma \lambda .
\end{equation}%
For the above line element the vector potential can be written as
\begin{eqnarray}
A_{t} &=&-\gamma h(r), \\
A_{r} &=&0, \\
A_{\theta } &=&\frac{\omega }{\alpha ^{2}}%
h(r), \\
A_{z} &=&0,
\end{eqnarray}%
where $h(r)$ is an arbitrary function of $r.$

\section{Scalar particles from rotating black strings}

To model the scalar tunneling from uncharged rotating black string
we use the Klein-Gordon equation for a scalar field $\Psi $ given by

\begin{equation}
g^{\mu \upsilon }\partial _{\mu }\partial _{\upsilon }\Psi -\frac{m^{2}}{%
\hslash ^{2}}\Psi =0.\   \label{95}
\end{equation}%
We apply the WKB approximation and assume an ansatz of the form

\begin{equation}
\Psi (t,r,\theta ,z)=e^{\left( \frac{i}{\hslash }I(t,r,\theta
,z)+I_{1}(t,r,\theta ,z)+O(\hslash )\right) },  \label{96}
\end{equation}%
where $I$ is the classical action of the trajectory. Now by using
Eq. (\ref{96}) in Eq. (\ref{95}) in leading order of $\hslash $ and
dividing by the exponential term and multiplying by $\hslash ^{2},$
we
get%
\begin{equation}
g^{tt}(\partial _{t}I)^{2}+g^{rr}(\partial _{r}I)^{2}+2g^{t\theta
}\partial _{t}I\partial _{\theta }I+g^{\theta \theta }(\partial
_{\theta }I)^{2}+g^{zz}(\partial _{z}I)^{2}+m^{2}=0.  \label{103}
\end{equation}%
The black string admits three Killing vectors $<\partial _{t},
\partial _{\theta }, \partial _{z}>$.
The existence of these symmetries implies that we can assume a solution for
Eq. (\ref{103}), in the form
\begin{equation}
I(t,r,\theta ,z)=-Et+W(r)+J_{1}\theta +J_{2}z+K,  \label{104}
\end{equation}%
where $E,$ $J_{1},$ $K$ and $J_{2}$ are constants and, further, we
consider the radial trajectories only. Substituting Eq. (\ref{104})
in Eq. (\ref{103}) and solving for $W(r)$, we get

\begin{equation}
W_{\pm }(r)=\pm \int \frac{\sqrt{E^{2}+n(r)(J_{1})^{2}-2N^{\phi
}EJ_{1}-w(r)(J_{2})^{2}-N^{0^{2}}m^{2}}}{\left( \gamma
^{2}-\frac{\omega ^{2}}{\alpha ^{2}}\right) g(r)(r/R)}dr, \label{aq}
\end{equation} 
where 
\begin{eqnarray}
n(r) &=&N^{\phi ^{2}}-\frac{N^{0^{2}}}{R^{2}},  \nonumber  \\
w(r) &=&-\frac{N^{0^{2}}}{\alpha ^{2}r^{2}},  \nonumber  \\
g(r) &=&\alpha ^{2}r^{2}-\frac{b}{\alpha r}.  \label{ab}
\end{eqnarray} 
Noting that at $r=r_{+}$ we have a simple pole and, therefore, by
using the residue theory for semi circle the integral yields
\begin{equation}
W_{\pm }(r)=\pm \pi i\frac{\sqrt{E^{2}+n(r_{+})(J_{1})^{2}-2N^{\phi
}(r_{+})EJ_{1}-w(r_{+})(J_{2})^{2}}}{\left( \gamma ^{2}-\frac{\omega ^{2}}{%
\alpha ^{2}}\right) g^{\prime }(r_{+})(r_{+}/R(r_{+}))}, \label{122}
\end{equation}
where 
\begin{eqnarray*}
g^{\prime }(r_{+}) &=&2\alpha ^{2}r_{+}+\frac{b}{\alpha r_{+}^{2}}, \\
R(r_{+}) &=&\gamma r_{+}.
\end{eqnarray*}%
From Eq. (\ref{122}) we see that
\begin{equation}
Im W_{\pm }(r)=\pm \pi \frac{\gamma \sqrt{
E^{2}+n(r_{+})(J_{1})^{2}-2N^{\phi }(r_{+})EJ_{1}-w(r_{+})(J_{2})^{2}}}{
\left( \gamma ^{2}-\frac{\omega ^{2}}{\alpha ^{2}}\right) g^{\prime }(r_{+})}
,  \label{123}
\end{equation}%
Now, the probabilities of crossing the horizon from inside to
outside and outside to inside are given by \cite{5, 6}

\begin{eqnarray}
P_{emission}  &\varpropto &\exp \left( \frac{-2}{\hbar }Im I\right)
=\exp \left[ \frac{-2 }{\hbar }\left(Im W_{+}+Im K \right)\right] ,
\label{1b} \\
P_{absorption} &\varpropto &\exp \left( \frac{-2}{\hbar }Im I\right)
=\exp \left[ \frac{-2 }{\hbar }\left(Im W_{-}+Im K \right)\right] . 
\end{eqnarray}
As the probability of any incoming particles crossing the horizons
have a 100\% chance of entering the black hole, therefore, it is
necessary to set

\begin{equation}
Im K=-Im W_{-},   \label{131}
\end{equation}%
in the above equation. From Eq. (\ref{122}), we get%
\begin{equation}
W_{+}=-W_{-}.  \label{132}
\end{equation}%
This means that the probability of a particle tunneling from inside
to outside the horizon is
\begin{equation}
\Gamma =\exp \left( -\frac{4}{\hslash }Im W_{+}\right) . \label{133}
\end{equation}%
From Eq. (\ref{123}), putting the value of $Im W_{+}$ in Eq. (\ref{133}%
), we get 

\begin{equation}
\Gamma = \exp \left[\left(\frac{ -4\pi \gamma}{\left( \gamma ^{2}-\frac{\omega ^{2}}{\alpha ^{2}}\right) g^{\prime }(r_{+})}\right) \left( \sqrt{
E^{2}+n(r_{+})(J_{1})^{2}-2N^{\phi }(r_{+})EJ_{1}-w(r_{+})(J_{2})^{2}}\right)\right] .  \label{134}
\end{equation} 
This is the probability of an outgoing scalar particle from the
event horizon of rotating black string. We note that this depends upon the mass of the black hole. If we compare this with the Boltzmann factor \cite{5, 6, DJ1} we see that the term in the first bracket of the exponential function is the inverse of 
Hawking temperature for uncharged rotating black string, that is

\begin{equation}
T_{H}=\frac{\left( \gamma ^{2}-\frac{\omega ^{2}}{\alpha
^{2}}\right) g^{\prime }(r_{+})}{4\pi \gamma }  \label{135}
\end{equation}%
or%
\begin{equation}
T_{H}=\frac{1}{4\pi \gamma }\left( \gamma ^{2}-\frac{\omega
^{2}}{\alpha ^{2} }\right) \left( 2\alpha ^{2}r_{+}+\frac{b}{\alpha
r_{+}^{2}}\right) .
\end{equation}%
By using Eqs. (\ref{2aa}) and (\ref{2aaa}), we note that $\gamma
^{2}-(\omega ^{2}/\alpha ^{2})=1$. So the temperature for uncharged rotating black string becomes
\begin{equation}
T_{H}=\frac{1}{4\pi \gamma }\left( 2\alpha ^{2}r_{+}+\frac{b}{\alpha
r_{+}^{2}}\right) .
\end{equation}

\section{Scalar particles from charged rotating black strings}

To study the contribution of scalar particles towards Hawking
radiation from charged rotating black strings, we use the charged
Klein-Gordon equation for scalar field $ \Psi \left( t,r,\theta
,z\right)$
\begin{equation}
\frac{1}{\sqrt{-g}}\left( \partial _{\mu }-\frac{iq}{\hbar }A_{\mu
}\right) \left( \sqrt{-g}g^{\mu \upsilon }(\partial _{\nu
}-\frac{iq}{\hbar }A_{\nu })\Psi \right) -\frac{m^{2}}{\hslash
^{2}}\Psi =0.  \label{144}
\end{equation}
Following a procedure similar to that of the previous section, we
let
\begin{equation}
\Psi (t,r,\theta ,z)=e^{\left( \frac{i}{\hslash }I(t,r,\theta
,z)+I_{1}(t,r,\theta ,z)+O(\hslash )\right) }.  \label{145}
\end{equation}%
Taking summation on $\mu $ and $\nu $ in Eq. (\ref{144}) and using
Eq. (\ref {145}) in leading order of $\hbar ,$ we get the
differential equation of the form
\begin{eqnarray}
&&g^{tt}(\partial _{t}I-qA_{t})^{2}+g^{rr}(\partial
_{r}I)^{2}+2g^{t\theta }(\partial _{t}I\partial _{\theta
}I-2qA_{t}\partial _{\theta
}I+q^{2}A_{t}A_{\theta })  \nonumber  \\
&&\left. +g^{\theta \theta }(\partial _{\theta }I-qA_{\theta
})^{2}+g^{zz}(\partial _{z}I)^{2}+m^{2}=0.\right.   \label{145a}
\end{eqnarray}%
By assuming a solution of the form in Eq. (\ref{104}) for the above
and evaluating for $W(r)$ gives

\begin{eqnarray*}
W_{\pm }(r) =\pm \int \frac{dr}{\left( \gamma ^{2}-\frac{\omega
^{2}}{\alpha ^{2}}\right) g(r)(r/R)}
\left[(E+qA_{t})^{2}+n(r)(J_{1}-qA_{\theta })^{2}  \right.
\nonumber
\end{eqnarray*}

\begin{equation}
\left. -2N^{\phi }(EJ_{1}+2qA_{t}J_{1}-q^{2}A_{t}A_{\theta
})-w(r)(J_{2})^{2}-N^{0^{2}}m^{2}\right]^{1/2} ,  \label{145b}
\end{equation}%
where%
\begin{eqnarray*}
n(r) &=&N^{\phi ^{2}}-\frac{N^{0^{2}}}{R^{2}}, \\
w(r) &=&-\frac{N^{0^{2}}}{\alpha ^{2}r^{2}}, \\
g(r) &=&\alpha ^{2}r^{2}-\frac{b}{\alpha r}+\frac{c^{2}}{\alpha
^{2}r^{2}}.
\end{eqnarray*}%
Here, we have a simple pole at $r=r_{+}$, and thus, by the residue
theory we evaluate the integral as
\begin{eqnarray*}
W_{\pm }(r) &=&\pm \frac{\pi i\gamma }{\left( \gamma ^{2}-\frac{\omega ^{2}}{%
\alpha ^{2}}\right) g^{\prime }(r_{+})}
\left[(E+qA_{t})^{2}+n(r_{+})(J_{1}-qA_{\theta })^{2} \right.
\end{eqnarray*}
\begin{equation}
\left. -2N^{\phi }(r_{+})(EJ_{1}+2qA_{t}J_{1}-q^{2}A_{t}A_{\theta
})\right]^{1/2},
\end{equation}
where%
\begin{equation}
g^{\prime }(r_{+})=2\alpha ^{2}r_{+}+\frac{b}{\alpha r_{+}^{2}}-\frac{2c^{2}%
}{\alpha ^{2}r_{+}^{3}}.
\end{equation}
This implies that 
\begin{eqnarray*}
Im W_{\pm }(r) &=&\pm \frac{\pi \gamma }{\left( \gamma ^{2}-\frac{%
\omega ^{2}}{\alpha ^{2}}\right) g^{\prime
}(r_{+})}\left[(E+qA_{t})^{2}+n(r_{+})(J_{1}-qA_{\theta })^{2}
\right.
\end{eqnarray*}
\begin{equation}
\left. -2N^{\phi }(r_{+})(EJ_{1}+2qA_{t}J_{1}-q^{2}A_{t}A_{\theta
})\right]^{1/2}.
\end{equation}
Thus the probability of a particle tunneling from inside to outside
the horizon as given by Eq. (\ref{133}) on substituting the value of
$Im W_{+}$ from the above equation takes the form

\begin{eqnarray*}
\Gamma  =\exp \left[ \left(
\frac{-4\pi \gamma }{\left( \gamma ^{2}-\frac{\omega ^{2}}{\alpha ^{2}}%
\right) g^{\prime }(r_{+})}\right) \left( \left[
(E+qA_{t})^{2}+n(r_{+})(J_{1}-qA_{\theta })^{2} \right. \right.  \right. 
\end{eqnarray*}
\begin{equation}
\left. \left. \left.  -2N^{\phi 
}(r_{+})(EJ_{1}+2qA_{t}J_{1}-q^{2}A_{t}A_{\theta })\right]^{1/2}
\right) \right] .
\end{equation} 
This depends upon the charge of the particle and not its mass. Again, comparison with the Boltzmann factor \cite{5, 6, 11a} shows that the term in the first parentheses gives the inverse of Hawking temperature for rotating charged black string. Thus 

\begin{equation}
T_{H}=\frac{1}{4\pi }\frac{\left( \gamma ^{2}-\frac{\omega ^{2}}{\alpha ^{2}}
\right) g^{\prime }(r_{+})}{\gamma },
\end{equation}%
or

\begin{equation}
T_{H}=\frac{1}{4\pi \gamma }\left( 2\alpha ^{2}r_{+}+\frac{b}{\alpha
r_{+}^{2}}-\frac{2c^{2}}{\alpha ^{2}r_{+}^{3}}\right) ,\label{ht1}
\end{equation}
where we have used $\gamma ^{2}-(\omega ^{2}/\alpha ^{2})=1$. This
formula is consistent with the literature \cite{15, 20}.

\section{Conclusion}

Hawking radiations from black holes comprise the whole range of
spectrum of particles including fermions, bosons, gravitinos etc. In
particular, emission of scalar fields has been studied for various
spherically symmetric black holes. In this paper we have extended
this analysis to cylindrically symmetric and rotating black
cinfigurations. Using the so-called Hamilton-Jacobi method and the
complex path integration technique, we have solved the Klien-Gordon
equation in the background of charged rotating black strings.
Employing WKB approximation we have found the tunneling
probabilities of uncharged and charged particles crossing the event
horizon. An important consequence of this procedure is that we get
the correct value of Hawking temperature. The temperature for
charged rotating black strings is given in Eq. (\ref{ht1}). If we
put angular momentum equal to zero, the temperature reduces to that
for the charged non rotating black strings \cite{14a}. Taking charge
$Q$ to be zero gives temperature for the uncharged case as

\begin{equation} T=\frac{1}{2\pi }\left(
\alpha ^{2}r_{+}+\frac{2M}{\alpha r_{+}^{2}}\right) . \label{5.6}
\end{equation}

\acknowledgments

Useful discussion with Douglas Singleton is acknowledged.

\end{document}